\documentclass{nwkltrs}

\sidecaptionvpos{figure}{c}

\begin{document}

\title{Endogenous drivers of gender disparity in online dating}
\headertitle{Endogenous drivers of gender disparity in online dating}

\author[1,2]{Huaiyu Tan}
\author[1]{Chenyu Li}
\author[3,4]{Taha Yasseri}
\author[2]{Lei Shi}
\author[1,5]{Petter Holme}

\affil[1]{Department of Computer Science, Aalto University, Espoo, Finland}
\affil[2]{School of Statistics and Mathematics, Yunnan University of Finance and Economics, China}
\affil[3]{School of Social Sciences and Philosophy, Trinity College Dublin, Ireland}
\affil[4]{Faculty of Arts and Humanities, Technological University Dublin, Ireland}
\affil[5]{Center for Computational Social Science, Kobe University, Kobe, Japan}

\twocolumn[%
\begin{@twocolumnfalse}
  \vspace{-15mm}
\maketitle
\begin{abstract}
\noindent In its early days, online dating was heralded as a great equalizer, removing biases built into heterosexual courtship. However, as repeatedly observed, that prophecy was never fulfilled, and some biases have even become exacerbated. In this paper, we identify a general endogenous mechanism that can amplify a small initial gender gap in first-contact rates of heterosexual dating, even when women and men follow identical behavioral rules. This mechanism relies on assumptions about the participants' expectations of new contacts and their time constraints. We formulate this symmetry-breaking mechanism as a system of differential equations and analyze its fixed points and their stability. The mechanism explains the amplification of a gap, not its origin or direction, which may stem from social norms, preferences, or platform design.
\end{abstract}
\vspace{5mm}
\end{@twocolumnfalse}]
\thispagestyle{empty}

\section{Introduction}

Online dating is a platform-mediated phenomenon that directly or indirectly affects a large fraction of humanity~\cite{degim_online_2015,rosenfeld_disintermediating_2019}. Moreover, it is a rich source of general insights into individual and organizational behavior~\cite{rudder2014dataclysm}. Although platforms supporting online heterosexual dating are surprisingly diversified into niches~\cite{degim_online_2015}, they share functionalities and constraints that enable generalization. For example, behavioral norms and customs can be learned through interactions with others on the site and thus spread from one part of the platform's social network to the rest over time~\cite{centola2018how,holme_edling_liljeros,bruch_aspirational_2018}.

Most of the empirical quantitative literature on online dating has taken a data-science approach, aiming to infer mechanisms underlying interaction dynamics from data~\cite{stoicescu_globalized_2019,zhang2016swipe,hitsch2010matching}. Some studies have quantified the implicit attractiveness and its effects on mate choice~\cite{bruch_aspirational_2018,gg_ph}. Other studies have used a discrete-choice approach, attempting to reverse-engineer people's logic in partner choice~\cite{bruch_extracting_2016,lewis}, or taken a comparative angle on the problem by comparing online dating across the world~\cite{stoicescu_globalized_2019}. These studies are mainly about \emph{whom} people contact and reply to---preferences over partner characteristics~\cite{fisman2006gender,hitsch2010matching}, desirability hierarchies~\cite{bruch_aspirational_2018}, and multistage screening~\cite{bruch_extracting_2016}. However, our focus is \emph{how much} each side initiates, given how much the other side initiates. We therefore leave preferences, screening, and reciprocation out of the model. Rather than treating them as competing explanations, we regard them as possible sources of a small initial asymmetry whose subsequent amplification is what our model describes. There are, however, few papers that take the approach of the present one---using mathematical (or sociophysics~\cite{JUSUP20221}) modeling to understand the mechanisms behind field observations.

Mathematical models that do exist address different questions. Two-sided matching theory~\cite{gale1962college} and matching platform models describe who ends up matched with whom, and how platform design affects welfare and competition, given agents' preferences~\cite{halaburda2018competing,kanoria2021facilitating}. Notably, Kanoria and Saban~\cite{kanoria2021facilitating} show that restricting which side may initiate can improve equality. Network models have been used to study how online dating changes the structure of society~\cite{ortega2017strength}, and general network-growth mechanisms such as preferential attachment describe the emergence of activity hierarchies~\cite{barabasi:book}. None of these models the feedback between the initiation activity of the two sides, which is the focus of the present paper.

One gendered moment of heterosexual courtship is who makes the first move. In the Western and Asian cultural spheres, before online dating, experiments show that such conventions shape gendered behavior---when women rather than men are assigned to approach in speed dating, the usual sex difference in selectivity disappears~\cite{finkel2009arbitrary}. Online dating is not strictly anonymous---gender, photographs and other profile information are visible from the outset---but first contacts are made in private, at a distance and without onlookers. Such settings can lower the pressure to follow conventional scripts~\cite{suler2004online}. In an early computer-dating system, a notable minority of women contacted many men, which the authors attributed to the system's safety and relative anonymity helping them depart from traditional sex-role norms~\cite{scharlott1995overcoming}. There was thus no \textit{a priori} reason for a norm of male initiation to persist online.
 
However, over time, this early sentiment that online dating services would promote gender equality~\cite{degim_online_2015} shifted to a recognition of widening gender gaps~\cite{kreager2014where,rosenfeld_disintermediating_2019,potarca2021online,dinh_computational_2020}. This seems contrary to the fact that general indicators of gender disparity, after all, show gradually narrowing gaps~\cite{gendergap}. An alternative explanation is that this phenomenon does not reflect societal change but rather is endogenous to the particular dating sites. In this paper, we propose a minimal model for such a mechanism in which the feedback between the ego's and alter's behavior destabilizes gender parity. The mechanism does not create a gender gap from nothing---it amplifies a small initial asymmetry, whatever its origin, even when women and men follow the same behavioral rule.

For the sake of our main argument, we assume that the two populations (women and men) are equal in size ($N/2$ individuals each). We further assume that each user has a preferred \emph{contact load}: the number of new first contacts per unit time---sent and received---that they are comfortable handling. Naturally, too few contacts are undesired, but even on efficient platforms, incoming contacts cost attention; beyond some point, additional contacts are worth less than they cost, consistent with the low reply rates among users receiving many, as observed in Ref.~\cite{bruch_aspirational_2018}. This equals an intermediate optimal contact load.

We now formulate this situation as a differential equation model of the invitation rates for the average member. The intuition behind the differential equations combines the idea that individuals have an ideal stream of new contacts and respond to the expected rate of incoming contacts. We will consider two cases. First, for well-mixed populations, we will design equations using mean-field dynamics and perform stability analysis of the modeled dynamical system. Second, we will formulate a corresponding model in which the system can be understood as a (bipartite) network of opportunities, represented by an adjacency matrix $\mathbf{A}=\{A_{ij}\}_1^N$. Such a network could be a practical consequence of the platform's algorithms controlling how others' profiles are presented to a user.

For the remainder of the paper, we will describe the model in detail and analyze it, both analytically and through numerical integration.

\section{Model}

\subsection{Variables and parameters}

We label the two populations $x$ and $y$; each has $N/2$ members. In the well-mixed (mean-field) model, $r_x$ is the mean rate, per member of $x$, at which first-contact invitations are \emph{sent} to members of $y$, normalized by the preferred contact load; $r_y$ is defined in the same way for $y$. Thus, $r_x$ is an outgoing rate, neither the incoming rate nor the sum of the two. Because the populations are equally large, every invitation sent by $x$ is received by $y$, so the mean incoming rate for a member of $y$ is $r_x$, and for a member of $x$ is $r_y$. A member of either group therefore handles the contact load $r_x+r_y$. In the networked model, each node $i$ has its own initiation rate $r_i$.

The asymmetry factor $\beta\geq 1$ measures how strongly members of $x$ respond to incoming invitations, relative to members of $y$; we choose the labels so that $x$ is the more responsive group. For $\beta=1$, the two groups follow identical rules, which is the case we use to isolate endogenous effects. Before rescaling, $0\leq s\leq 1$ is the expected rate of incoming first contacts, and $0<c\leq 1$ is the sensitivity to received invitations (larger $c$ means a stronger response to changes in the incoming rate).

\subsection{Well-mixed population}

The following system of equations is a formalization of the observations in the previous paragraph: 
\begin{equation} \label{eq:raw} 
 \begin{dcases}
 \dfrac{dr_x}{dt} = \big(1-r_x-r_y \big) \big|s - c \beta r_y \big|, \\
 \dfrac{dr_y}{dt} = \big(1-r_x-r_y \big) \big|s - c r_x \big|
 \end{dcases}
\end{equation}
Equation~\eqref{eq:raw} is a phenomenological adjustment rule, not an optimal response derived from a payoff function. Its two factors have separate roles. The capacity factor $1-r_x-r_y$ sets the direction of adjustment: users increase their initiation rate while their contact load is below the preferred level and decrease it above. It can be read as the linearization, around the preferred load, of the marginal net value of new contacts. The drive factor, $|s-c\beta r_y|$ for $x$ and $|s-cr_x|$ for $y$, sets the speed: the further the incoming stream is from expectations, in either direction, the faster users revise their activity. A user who receives fewer invitations than expected compensates by initiating more; a user who receives more than expected is encouraged to be more active, as long as the load permits. One can also imagine a signed variant, in which an excess of incoming invitations reduces one's own initiation, which we analyze in the next section; it amplifies initial differences at least as strongly. Without loss of generality we can rescale $t$ and $s$ by $c$, leading to
\begin{equation}\label{eq:mf}
 \begin{dcases}
 \dfrac{dr_x}{dt} = \big(1-r_x-r_y\big) \big|s - \beta r_y \big|, \\
 \dfrac{dr_y}{dt} = \big(1-r_x-r_y \big) \big|s - r_x \big|
 \end{dcases}
\end{equation}
where $s>0$ is the rescaled expected incoming rate, which we call the expectation parameter. After this rescaling, $s$ is not mathematically restricted to $[0,1]$; the figures below only display selected parameter windows needed to illustrate the regimes considered.

The model describes first-contact attempts, not reciprocated matches. Gender differences in reply probabilities are documented~\cite{bruch_aspirational_2018}, but we ask whether initiation disparities can be amplified without such differences as an input. Fixed reply probabilities can nevertheless be absorbed into the parameters. If members of $x$ and $y$ reply to an incoming invitation with probabilities $p_x,p_y>0$, and only replied-to contacts count against expectations, the drives in Eq.~\eqref{eq:raw} become $|s-c\beta p_x r_y|$ and $|s-cp_y r_x|$. This is Eq.~\eqref{eq:raw} with $c'=cp_y$ and $\beta'=\beta p_x/p_y$ (relabeling the populations, and rescaling $c$, if $\beta'<1$). A gender difference in reply rates is thus one possible source of $\beta\neq 1$. This reparameterization does not cover workloads that depend on acceptance, or acceptance that adapts to the number of approaches; such features add further feedback, whereas the present model isolates the feedback in initiation activity.

\section{Signed adjustment rule}\label{app:signed}

\begin{SCfigure*}[1.3]
 \includegraphics[width=1.2\linewidth]{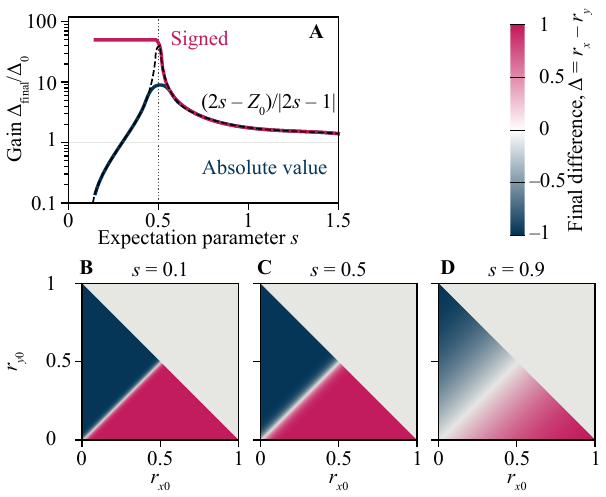}
 \caption{\textbf{Absolute-value versus signed adjustment ($\beta=1$).}
 (A) Final gap relative to the initial gap, for initial load $Z_0=0.2$ and initial gap $0.02$, as a function of $s$. Solid lines represent numerical integration of Eq.~\eqref{eq:mf} (blue) and Eq.~\eqref{eq:signed} (magenta). Dashed line shows Eqs.~\eqref{eq:gain} and \eqref{eq:gain_abs}. For $s\leq 1/2$, the signed rule ends in the maximally asymmetric state (gain $1/0.02=50$). Panels B, C, and D display the final gap $\Delta=r_x-r_y$ under the signed rule for initial conditions with $r_{x0}+r_{y0}<1$ for various $s$. Grey means outside the feasible domain. Cf.\ Fig.~\ref{fig:final}.}
 \label{fig:signed}
\end{SCfigure*}

To test whether amplification depends on the absolute value in Eq.~\eqref{eq:mf}, we replace it with a signed drive, so that users initiate more when they receive fewer invitations than expected and less when they receive more:
\begin{equation}\label{eq:signed}
\dot r_x=(1-Z)\,(s-\beta r_y),\qquad \dot r_y=(1-Z)\,(s-r_x),\qquad Z=r_x+r_y .
\end{equation}
Because the drive can now be negative, the feasible domain $r_x,r_y\geq 0$, $Z\leq 1$ must be imposed explicitly. For $\beta=1$, the gap $u=r_x-r_y$ and the load $Z$ obey
\begin{equation}
\dot u=(1-Z)\,u,\qquad \dot Z=(1-Z)\,(2s-Z)
\end{equation}
everywhere in the interior, not only near the symmetric point. The gap is therefore amplified for every $s$. For $s>1/2$, $Z$ increases to the saturation line, and $du/dZ=u/(2s-Z)$ integrates to
\begin{equation}\label{eq:gain}
\frac{u_\infty}{u_0}=\frac{2s-Z_0}{2s-1},
\end{equation}
which coincides with the absolute-value rule as long as both rates stay below $s$ (always the case for $s\geq 1$). For $s<1/2$, the load converges to $Z=2s$, the symmetric point $(s,s)$ is a saddle with eigenvalues $\pm(1-2s)$, and the gap grows until one group stops initiating altogether. The signed rule thus amplifies initial differences at least as strongly as in Eq.~\eqref{eq:mf}; the damping in Regime~I is a consequence of the absolute value (Fig.~\ref{fig:signed}). We nevertheless keep the absolute value in the main model: with the signed drive, $\dot Z>0$ whenever $Z>\max(1,2s)$, so activity exceeding the preferred load can grow without bound---for $s<1/2$, the saturation line repels trajectories from above---and the non-negativity of the rates has to be imposed by hand.

\subsection{Networked population}

For the networked case, we can generalize the model above to:
\begin{equation}
\label{eq:nwk}
\dfrac{dr_i}{dt} = \left \{
\begin{array}{ll} \big(1-r_i-R_i\big) \big| s - \beta R_i \big|, & \text{ if $i$ is $x$} \\
 \big(1-r_i-R_i \big) \big|s - R_i\big|, & \text{ if $i$ is $y$}\end{array} \right.
\end{equation}
where $R_i$, the invitation rates of $i$, is:
\begin{equation}
 R_{i} = \sum_{j} A_{ij} r_j ,
\end{equation}
with $A_{ij}=1/k_i$ if $i$ and $j$ are linked and $A_{ij}=0$ otherwise,
where $k_i$ is the degree of $i$. This ensures that on the complete bipartite graph
Eq.~\eqref{eq:nwk} reduces to the mean-field model

Note that in this case, the rates are for individual nodes, not averages over the two sub-populations.

\section{Results}

We proceed with our analysis of the model. First, we present the mean-field approximation as a dynamical system. Second, we discuss our numerical analysis of the networked case.

\subsection{Fixed point analysis}

To find the fixed points of the mean-field equations Eq.~\eqref{eq:mf}, let $\dot{r}_x = \dot{r}_y = 0$. The equations factor into a saturation factor $1 - r_x - r_y$ and a drive factor, yielding two conditions for a fixed point. From the saturation factor
\begin{equation}
 1 - r_x - r_y = 0 \quad \Rightarrow \quad r_x + r_y = 1
\end{equation}
This defines a saturation line of fixed points. A member of $x$ sends invitations at rate $r_x$ and receives them at rate $r_y$, so their contact load is $Z=r_x+r_y$; the same holds for a member of $y$. On the saturation line, every user handles exactly their preferred load. Initiation rates then stop changing, but invitations continue to flow at constant rates; reaching the line implies neither that users have found partners nor that their expectations of incoming contact are met. The saturation factor thus represents a common per-user contact budget, normalized to one, in which sent and received invitations count equally---not a platform-wide resource. In the networked model, the corresponding factor $1-r_i-R_i$ is individual. Below saturation, the common positive factor $1-Z$ only changes the speed along trajectories, not their directions. For the drive factor, we obtain:
\begin{equation}\left\{
\begin{array}{lll}
 s - \beta r_y = 0 & \Rightarrow & r_y = s/\beta \\
 s - r_x = 0 & \Rightarrow & r_x = s
\end{array}\right. .
\end{equation}
This yields a unique isolated fixed point in the interior of the phase space
\begin{equation}
 \big(r_x^*, r_y^*\big) = \left( s, \dfrac{s}{\beta} \right) .
\end{equation}

\begin{SCfigure*}[1.3]
 \includegraphics[width=0.8\linewidth]{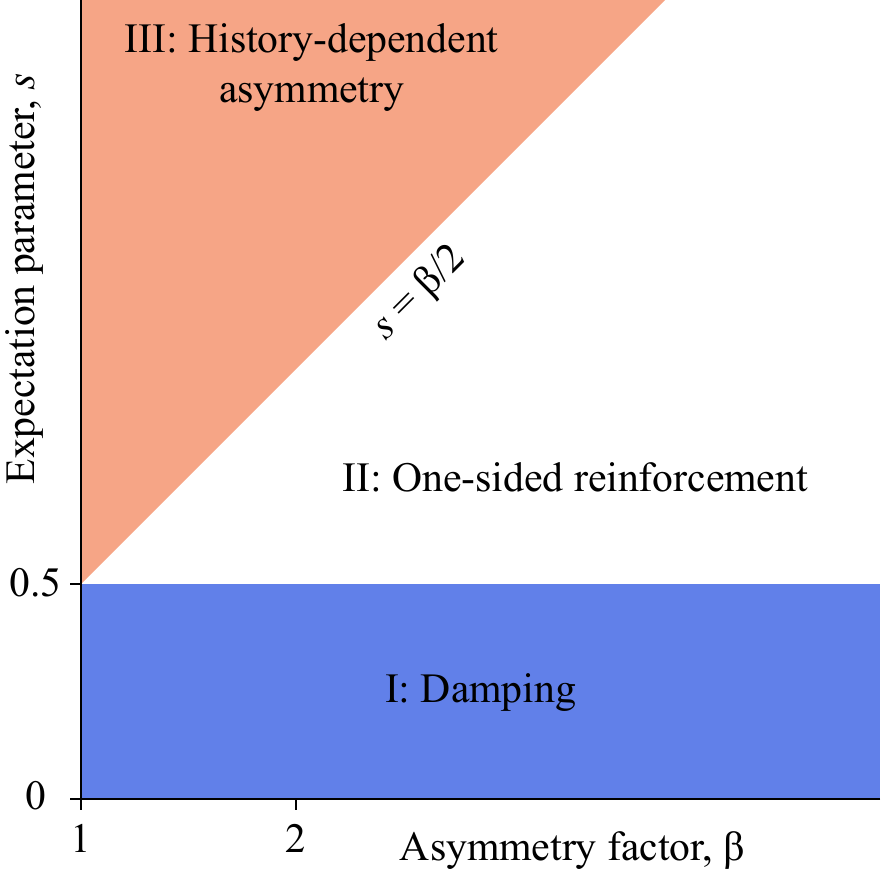}
 \caption{
\textbf{Phase diagram.} This figure shows the system's qualitative behavior before saturation for the continuous asymmetry factor ($\beta$, x-axis) and expectation parameter ($s$, y-axis) parameters. The boundary lines ($s=1/2$ and $s=\beta/2$) divide the space into three distinct dynamic regimes: The symmetric saturation, $x$ dominance, and history-dependent asymmetry. These boundaries should be read as changes in the feedback before the trajectory reaches the saturation line, rather than as motion along that line after saturation. Low overall user drive ($s<0.5$) tends to reduce the difference between the two rates, whereas higher expectations tend to amplify such differences before the final saturation point is reached.}
 \label{fig:phd}
\end{SCfigure*}

\begin{figure*}[htbp]
 \centering
 \includegraphics[width=0.8\linewidth]{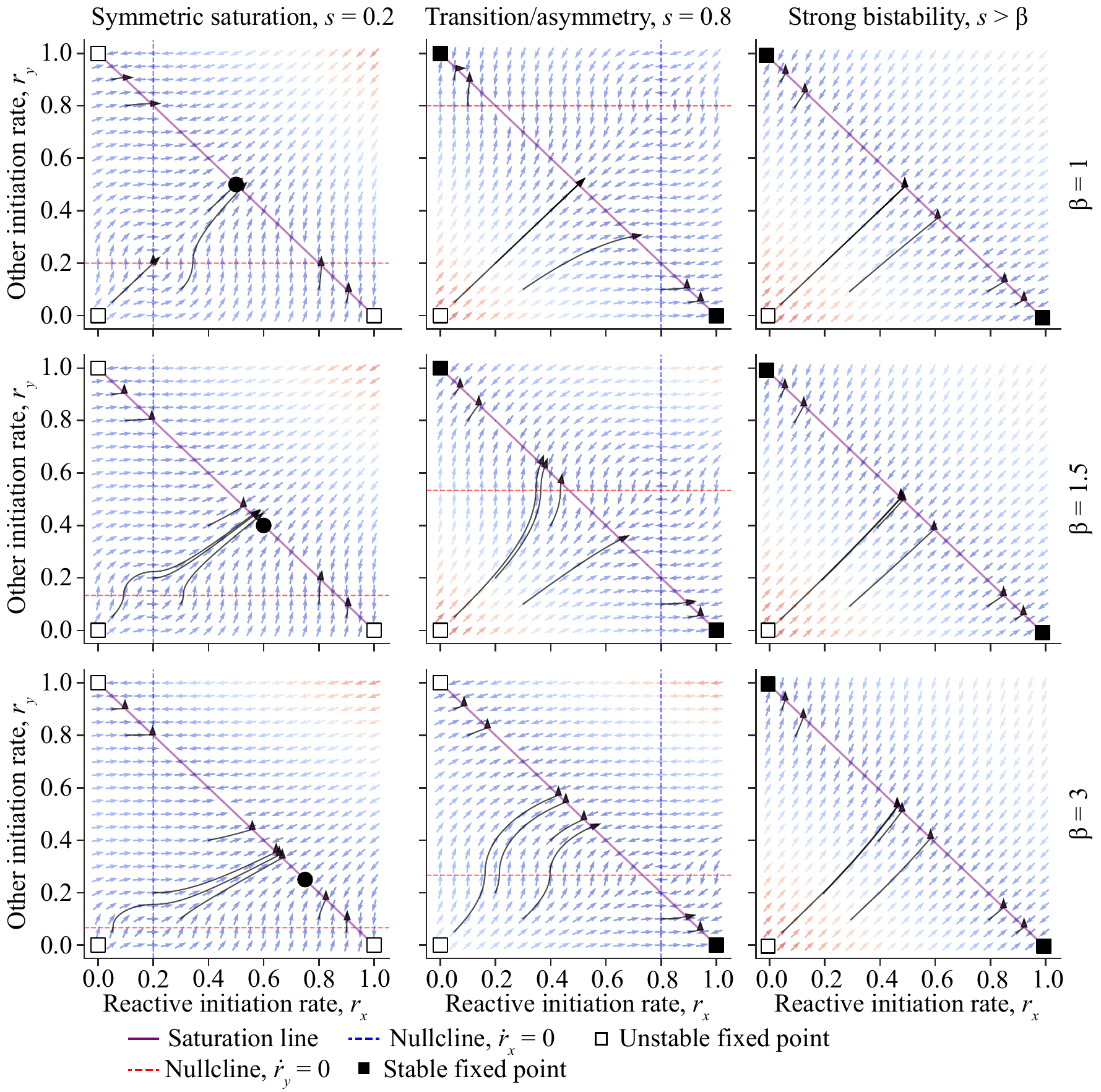}
 \caption{\textbf{Phase and stability analysis in the $(r_x, r_y)$ plane.} 
Vector fields (arrows) illustrate the gradient of the system's evolution, with color intensity indicating the magnitude of the time derivatives. The purple solid line represents the physical saturation constraint $r_x + r_y = 1$. Dashed lines indicate the nullclines for the $x$ population ($\dot{r}_x=0$) and $y$ ($\dot{r}_y=0$). Strictly speaking, the saturation line is also part of both nullclines; the dashed lines show only the drive-zero branches. Markers denote fixed points and limiting saturation states: points on the purple line are fixed once reached, because the common factor $1-r_x-r_y$ then vanishes. The rows correspond to $\beta=1$, $1.5$, and $3$ respectively. Below the line, the flow points towards the saturation line everywhere except at the drive-zero fixed point $(s,s/\beta)$; once on the line, the dynamics stop. The columns correspond to $s=0.2$ (Regime~I for all $\beta$), $s=0.8$ (Regime~III for $\beta=1$ and $1.5$, Regime~II for $\beta=3$) and $s=\ldots$ (Regime~III).
}
 \label{fig:phase_portraits}
\end{figure*}

\subsection{Stability analysis}

 Next, we address the stability of these fixed points. Since the system involves absolute-value functions, the vector field is not differentiable along the drive-zero lines, where the Jacobian is undefined. However, we can analyze the stability by examining the vector field in the regions defined by the saturation line. Consider the region inside the feasible domain where total activity is below capacity ($Z = r_x + r_y < 1$). In this region, $(1 - Z) > 0$. Since the absolute value function outputs non-negative values ($| \cdot | \ge 0$), we have
\begin{equation}
\frac{dr_x}{dt} \ge 0 \quad \text{and} \quad \frac{dr_y}{dt} \ge 0
\end{equation}
for all points inside the domain. Because the derivatives are generally non-negative, trajectories inside the domain ($Z < 1$) move monotonically towards the upper-right, apart from the drive-zero branches where one component may momentarily vanish. Unlike the linear model, in which the drive can become negative and cause a collapse to zero, the absolute-value formulation ensures that users continue to initiate activity until the capacity constraint is reached. The isolated drive-zero point $(s, s/\beta)$, when it lies inside the feasible domain, is a fixed point that is not an attractor. Except at this point, trajectories evolve towards the saturation line. In this sense, the saturation line attracts trajectories from the interior, although the dynamics stop once the line is reached.

The fixed points on the line $r_x+r_y=1$ attract trajectories from the interior, but every point on this line is already a fixed point of Eq.~\eqref{eq:mf}. Indeed, if $Z=r_x+r_y$, then $\dot Z=(1-Z)(D_x+D_y)$, where
\begin{equation}
D_x = |s - \beta r_y| \quad \text{and} \quad D_y = |s - r_x| .
\end{equation}
Thus, the comparison between $D_x$ and $D_y$ should not be interpreted as motion after the trajectory has reached the line. It only describes which component grows faster just before saturation and, therefore, which part of the saturation line is approached.

With this interpretation, the useful threshold near the endpoint $(0,1)$ is obtained from $|s|=|s-\beta|$, giving
\begin{equation}
s = \frac{\beta}{2} .
\end{equation}
Similarly, near the endpoint $(1,0)$, the threshold $|s|=|s-1|$ gives
\begin{equation}
 s = \frac{1}{2} .
\end{equation}
These conditions mark changes in the bias before saturation toward different parts of the saturation line, not asymptotic stability along the line itself.

When $\beta=1$, the system reduces to
\begin{equation}
\begin{cases}
\frac{dr_x}{dt} = (1 - r_x - r_y) |s - r_y| \\
\frac{dr_y}{dt} = (1 - r_x - r_y) |s - r_x|
\end{cases}
\label{eq:system_beta1}
\end{equation}
To understand the evolution of the gender disparity in this case, we define the gap variable $u = r_x - r_y$. The time derivative of this gap is
\begin{equation}
\frac{du}{dt} = \frac{dr_x}{dt} - \frac{dr_y}{dt} = (1 - Z) \Big( |s - r_y| - |s - r_x| \Big)
\label{eq:gap_derivative}
\end{equation}
where $Z = r_x + r_y$. Since $1 - Z > 0$ strictly in the interior of the feasible domain, the growth or decay of the gender gap $u$ depends entirely on the sign of $\big( |s - r_y| - |s - r_x| \big)$. Depending on the magnitude of the expectation parameter $s$, the system exhibits two fundamentally distinct feedback regimes.

When $s>1/2$, consider the dynamics near the symmetric saturation point $(1/2, 1/2)$. If the expectation parameter is high ($s > 1/2$), then in the vicinity of this point, both $r_y < s$ and $r_x < s$. Consequently, the absolute values can be dropped directly: $|s - r_y| = s - r_y$ and $|s - r_x| = s - r_x$. Substituting these into Equation \eqref{eq:gap_derivative} yields
\begin{equation}
\frac{du}{dt} = (1 - Z) \Big( (s - r_y) - (s - r_x) \Big) = (1 - Z) (r_x - r_y) = (1 - Z) u
\end{equation}
Since $1 - Z > 0$, the derivative $du/dt$ has the same sign as $u$. This implies that if the $x$ population initiates slightly more than $y$s ($u > 0$), their drive to initiate grows even faster than that of $y$s, pushing $u$ further in the positive direction. This is a local amplification of the gap before saturation. It does not imply flow along the saturation line, because $1-Z=0$ on that line. Thus, a high baseline expectation ($s>1/2$) can amplify small initial differences before saturation and can produce a strongly asymmetric final saturated state, despite the absence of any structural gender bias ($\beta=1$).

When $s < 1/2$, the expectation parameter is relatively low ($s < 1/2$), changing the dynamics as the system approaches saturation. Near the symmetric point $r_x=r_y=1/2$, both activity levels exceed the drive: $r_y > s$ and $r_x > s$. In this region, the absolute terms flip signs: $|s - r_y| = r_y - s$ and $|s - r_x| = r_x - s$. Substituting this into Equation \ref{eq:gap_derivative} yields
\begin{equation}
\frac{du}{dt} = (1 - Z) (r_y - r_x) = - (1 - Z) u
\end{equation}

Here, $du/dt$ and $u$ have opposite signs. This constitutes negative feedback: if one gender initiates more than the other, its effective drive decreases relative to the opposite gender, allowing the less active gender to catch up. Although a small gap $u$ might initially grow near the origin (where $r < s$), once the overall activity level crosses the threshold $s$, the negative feedback dominates. The gap is then reduced while the system is still below saturation. Since $1-Z\to 0$ as the saturation line is approached, this reduction need not force exact convergence to $(1/2,1/2)$; rather, the final point is typically closer to the symmetric part of the saturation line and depends on the preceding trajectory.

For small initial gaps, the final gap can be computed exactly. Writing $a=s-r_x$ and $b=s-r_y$, Eq.~\eqref{eq:system_beta1} conserves $b^2-a^2$ where $a$ and $b$ have the same sign and $a^2+b^2$ where they have opposite signs, while $1-Z$ only sets the speed. Following a trajectory that starts with both rates below $s$, load $Z_0$ and gap $u_0$ (small compared with $|2s-1|$) to the saturation line gives
\begin{equation}\label{eq:gain_abs}
\frac{u_\infty}{u_0}=\frac{2s-Z_0}{|2s-1|} .
\end{equation}
The gain exceeds one whenever $s>(1+Z_0)/4$---so a gap can grow on balance even in Regime~I, if it grows enough before the rates exceed $s$---and it diverges as $s\to 1/2$: amplification is strongest when users' expectations are close to what a balanced platform delivers.

\subsection{The mechanism in behavioral terms}

Suppose that, for historical or random reasons, one group $x$ initially sends slightly more first contacts than group $y$, both groups follow the same endogenous behavioral rule ($\beta=1$), and loads are below the preferred level, so that both groups are increasing their activity. Each $y$ individual then receives slightly more invitations than each $x$ individual. If both groups receive fewer invitations than they expect, $y$'s shortfall is the smaller one, so their initiation rate rises more slowly; $x$s, who receive fewer invitations, face the larger shortfall, and their initiation rate rises faster. The original difference widens and each side's initiation substitutes for the other's. If instead both groups receive more invitations than they expect, the $y$s surplus is the larger one. Under our rule, a larger deviation from expectations means faster adjustment, so the $y$ individual's initiation now rises faster, and the difference narrows. In both cases, both rates increase; what changes is which increases faster. Once the load reaches the preferred level, adjustment stops, and the difference accumulated up to that point is frozen in. Which of the two situations prevails near saturation depends on $s$---at the symmetric saturated state, each user receives exactly half of their contact load, so differences are amplified if users expect more than half of their contacts to be initiated by others ($s>1/2$) and damped if they expect less ($s<1/2$).

\subsection{Summary of dynamics}

Figure~\ref{fig:phd} summarizes three qualitative regimes. They are easiest to interpret using the rescaled expectation, the expected incoming rate measured in units of the preferred contact load $s=s_0/c$. At the symmetric saturated state, every user receives exactly half of their load. For group $y$, the relevant comparison is thus between $s$ and $1/2$; for group $x$, whose drive can be written $\beta\,|s/\beta-r_y|$, it is between $s/\beta$ and $1/2$. The boundaries $s=1/2$ and $s=\beta/2$ in Fig.~\ref{fig:phd} are exactly these two comparisons. They also determine whether a lopsided state reinforces itself---near such a state, the initiating group receives almost nothing and the other group almost everything, and the state is self-reinforcing if the under-contacted initiators adjust faster than the over-contacted others.
\begin{description}
 \item[Regime I (damping)] $s<1/2$. Here, both groups expect less than half of their load to come from incoming invitations. Near a balanced state, both are over-contacted; the less active group, which receives more, adjusts faster, and gaps shrink as the load approaches saturation. Lopsided states are not self-reinforcing. The final point still depends on the earlier trajectory, and a gap that grew early on is not necessarily removed.
 \item[Regime II (one-sided reinforcement)] When $1/2<s<\beta/2$, group $y$ expects more than half of its load from incoming invitations, whereas group $x$ expects less. Only lopsided states in which $x$ does the initiating are self-reinforcing.
 \item[Regime III (history-dependent asymmetry)] $s>\beta/2$. Here, both groups expect to be approached more than they approach. These expectations cannot both be met at a balanced state, and lopsided states led by either group are self-reinforcing; which group is approached depends on which one initiates more early on, much like a convention about who makes the first move~\cite{finkel2009arbitrary}. For $\beta=1$, the symmetric trajectory is unstable.
\end{description}
In all three regimes, the line $r_x+r_y=1$ itself consists of fixed points; the regimes describe how trajectories approach this line from the interior, and ``dominance'' refers to a higher outgoing invitation rate, not to greater success in finding partners. For $\beta>1$, the thresholds characterize the ends of the saturation line rather than the typical endpoint of a trajectory: starting from low, equal rates, the more responsive group $x$ ends ahead for small $s$, whereas for larger $s$ the less responsive group $y$ tends to end ahead (Fig.~\ref{fig:phase_portraits}), because $x$'s shortfall closes first once invitations start to arrive. Figure~\ref{fig:phase_portraits} shows the phase portraits, including vector fields, nullclines, sample trajectories, and fixed points for three values of the $\beta$ parameter.

\begin{figure*}[htbp]
 \centering
 \includegraphics[width=0.85\linewidth]{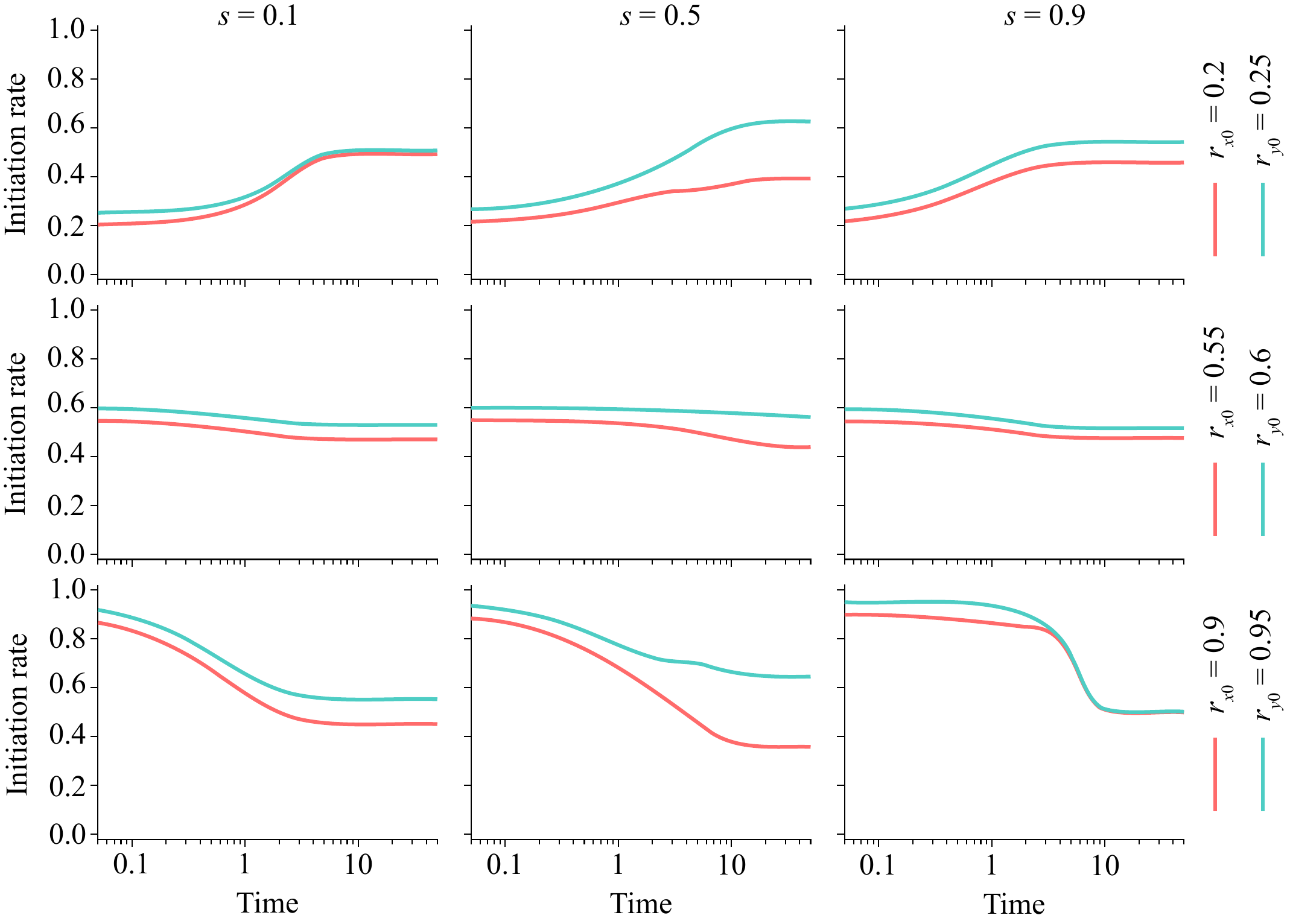}
 \caption{\textbf{Time evolution of initiation rates.} This figure shows the temporal evolution of $x$ and $y$ initiation rates on a logarithmic time scale. The $3 \times 3$ panels explore varying expectation parameters ($s=0.1$, $0.5$, $0.9$) across columns and different initial conditions across rows, assuming no exogenous forces affect the initiation rates (i.e., the gender balance parameter is balanced, $\beta=1$).}
 \label{fig:temporal_dynamics_again}
\end{figure*}

\begin{figure*}[htbp]
 \centering
 \includegraphics[width=0.85\linewidth]{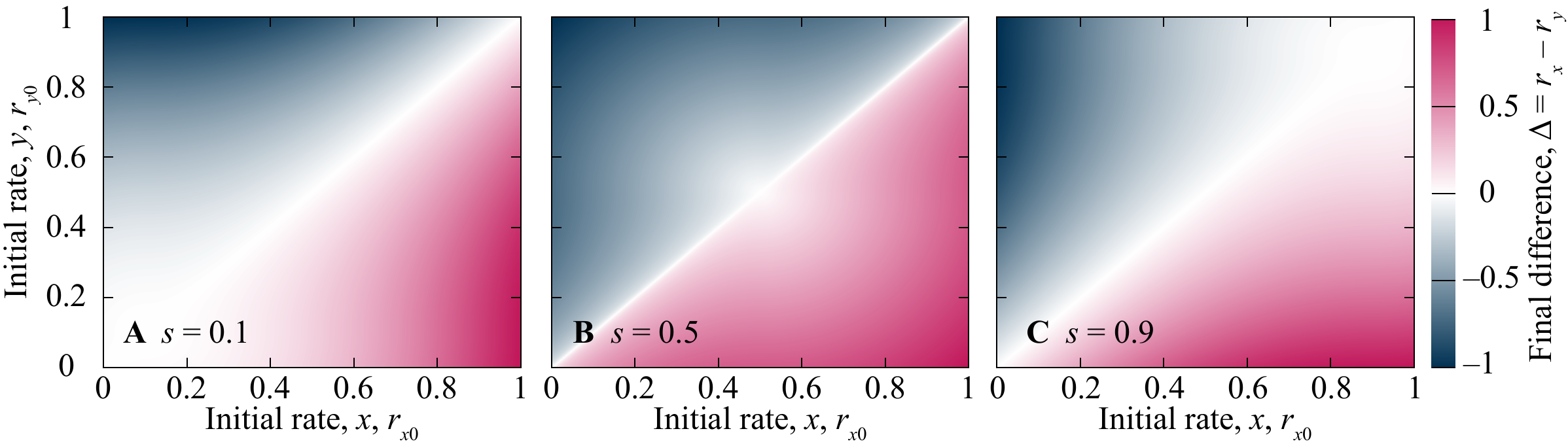}
 \caption{\textbf{Heatmap of final initiation rate differentiation.} This figure visualizes the final gender disparity ($\Delta = r_x - r_y$) over a $100 \times 100$ grid of the initial initiation rate parameter space. A diverging colormap represents $y$-dominant, perfectly symmetric, and $x$-dominant outcomes under symmetric sensitivity ($\beta = 1$).
}
 \label{fig:final}
\end{figure*}

\subsection{Dynamic simulations}
\label{sec:results_sim}

To corroborate the observations in the previous section, we integrate the system of equations in Eq.~\eqref{eq:mf} for representative parameter sets (see Fig.~\ref{fig:temporal_dynamics_again}). This figure explores the limiting case $\beta = 1$, in which $x$ and $y$ are identical. As expected from Figs.~\ref{fig:phd} and \ref{fig:phase_portraits}, the symmetry can break through amplification of small initial differences. In particular, for an intermediate expectation parameter ($s=0.5$) and off-center initial conditions ($r_0\approx 0.2$ and $0.9$), the differences in $r$ diverge significantly between the two groups. Thus, even in the absence of noise and without exogenous symmetry-breaking ($\beta>1$), the dynamics before saturation can leave a clear emergent gender disparity in the final saturated values of $r$. Figure~\ref{fig:final} generalizes Fig.~\ref{fig:temporal_dynamics_again}'s observations about the final deviation's $r_{x0}$- and $r_{y0}$-dependence. It confirms that for intermediate $s$-values, a larger fraction of the initial conditions leads the system to a given deviation $\Delta$.

\begin{figure*}[htbp]
 \centering
 \includegraphics[width=0.85\linewidth]{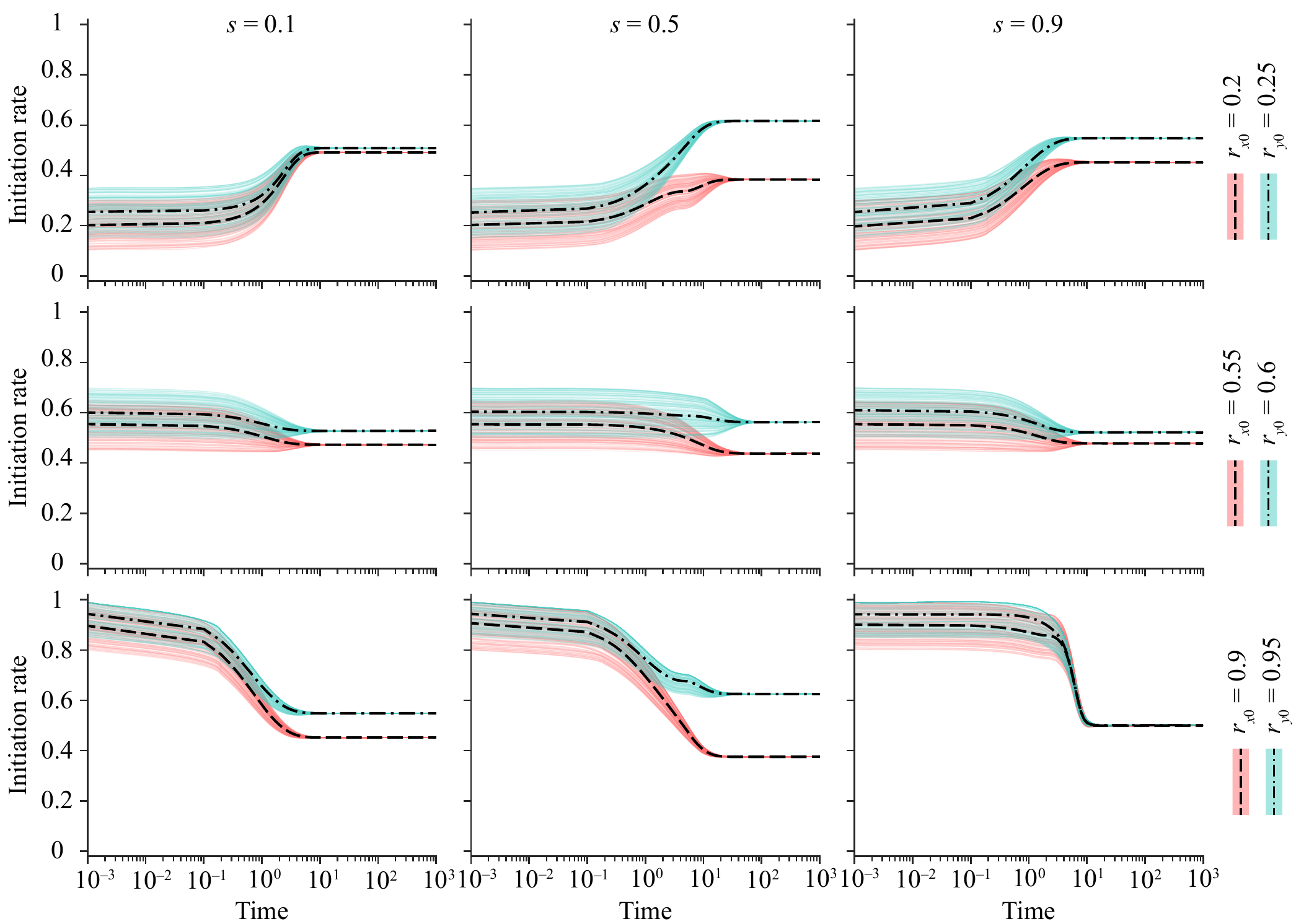}
 \caption{\textbf{Time evolution of initiation rates (individual and mean) on a complete network.} This figure traces the temporal dynamics of $N=5000$ individuals on a fully connected topology, plotting macroscopic gender means (black dashed lines) alongside individual trajectories (thin lines). We vary the expectation parameters while keeping $\beta=1$. 
}
 \label{fig:time_complete}
\end{figure*}

\begin{figure*}[htbp]
 \centering
 \includegraphics[width=0.85\linewidth]{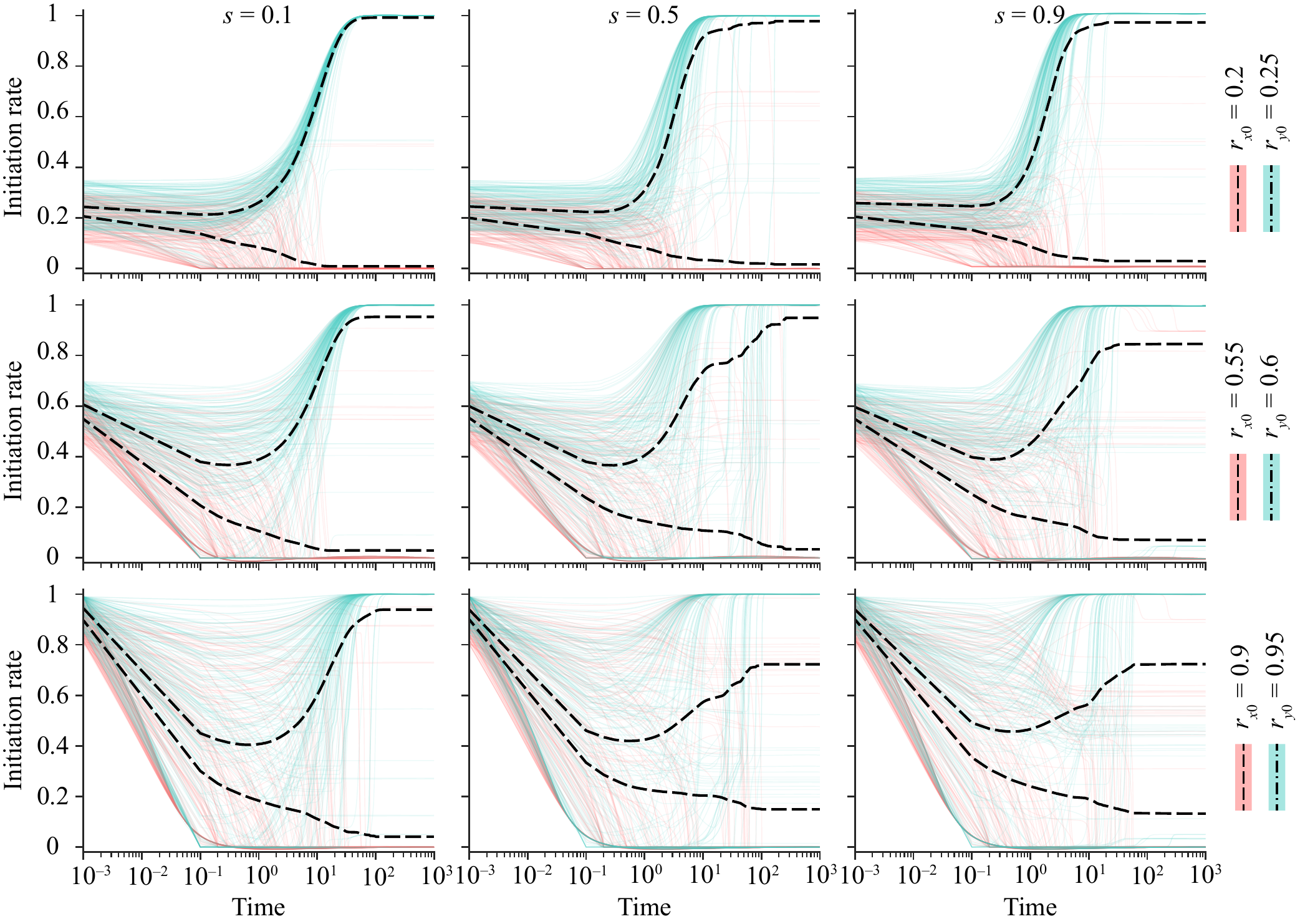}
 \caption{\textbf{Time evolution of initiation rates (individual and mean) on an Erd\H{o}s--R\'enyi random network.} Numerical integration of Eq.~\eqref{eq:nwk} for $N=5000$ users on a bipartite Erd\H{o}s–R\'enyi graph with average degree three, showing global means and individual trajectories under symmetric sensitivity ($\beta=1$).
}
 \label{fig:time_er}
\end{figure*}

\begin{figure*}[htbp]
 \centering
 \includegraphics[width=0.85\linewidth]{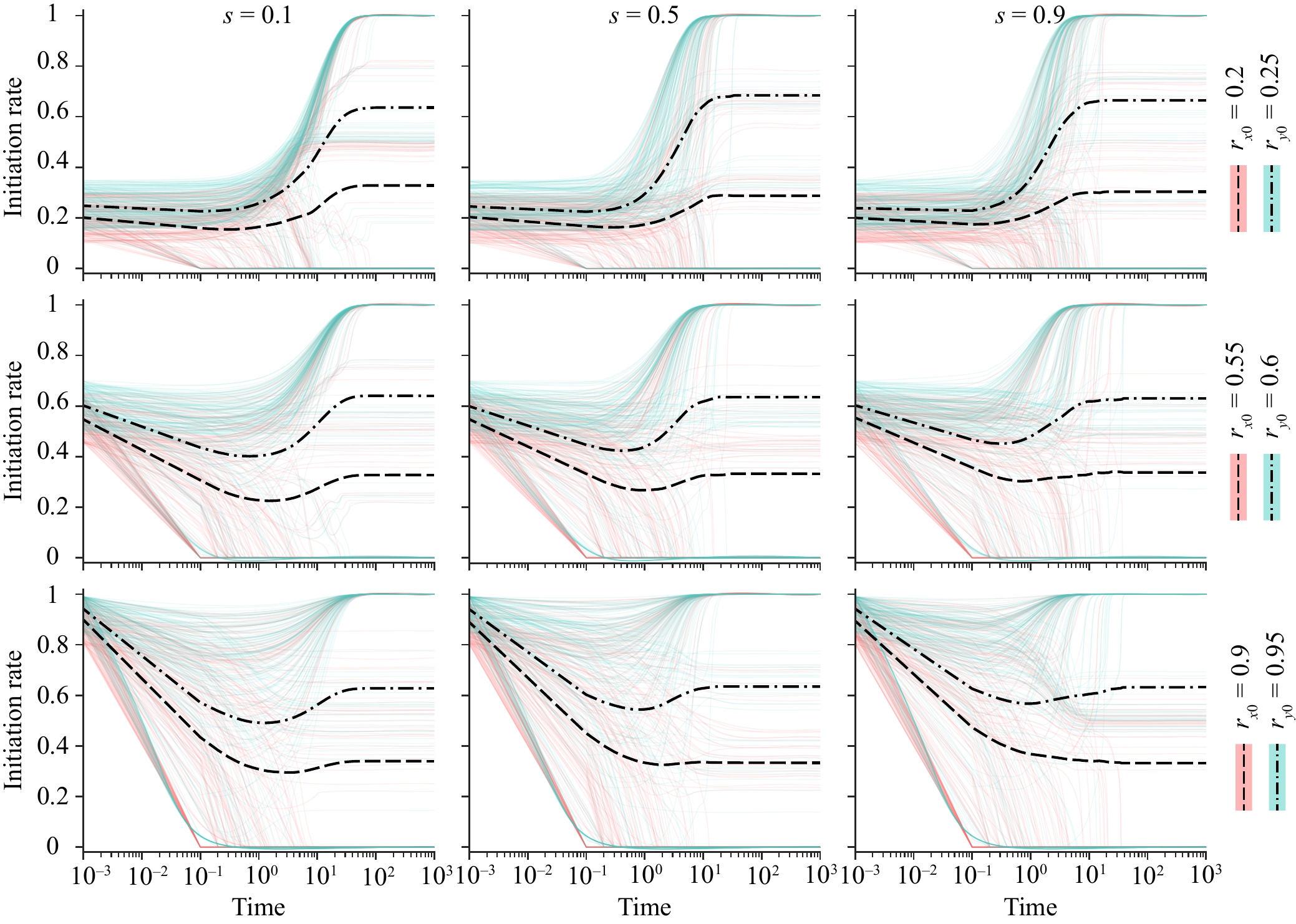}
 \caption{\textbf{Time evolution of initiation rates (individual runs and mean values) on a Barab\'asi--Albert scale-free network.} This figure illustrates dynamics for $N=5000$ users on the highly heterogeneous bipartite Barab\'asi--Albert network (average degree $3$). Other parameter values are also as for Fig.~\ref{fig:time_er}.
}
 \label{fig:time_ba}
\end{figure*}

\subsection{Networked interaction}

Next, we move to the networked versions of the model. We integrate Eq.~\eqref{eq:nwk} for different networks: fully connected graphs, and bipartite versions of the Erd\H{o}s-R\'enyi model (also known as random graphs) and the Barab\'asi-Albert model of networks with emergent power-law degree distributions~\cite{mejn:book,barabasi:book,guillaume-2004-bipartite}. 

Figure~\ref{fig:time_complete} shows the results for fully-connected graphs, confirming that it retains the behavior of the mean-field solution. Just like in Fig.~\ref{fig:temporal_dynamics_again}, intermediate $s$-values induce the largest deviations in the final initiation rates. We sample the initial initiation rates from a rectangular distribution with width $0.1$. The dynamics then force the individuals' trajectories to converge. This reflects how the Law of Large Numbers suppresses fluctuations in many forms of dynamics on networks, from neuronal dynamics~\cite{angu_pcb} to epidemic models~\cite{Duron2022}, where, typically, fluctuations in population averages scale like the inverse square root of the average degree.

In Fig.~\ref{fig:time_er}, we see that the corresponding plot for Erd\H{o}s-R\'enyi networks shows much less stable dynamics, with the disparity approaching its maximum throughout the parameter space. Indeed, it is very rare for individuals to reach an initiation rate that is neither the minimum (zero) nor the maximum (one). The dynamics is, furthermore, very dramatic---with plateaus and sudden changes---and worthy of a more detailed study in its own right.

Our final network topology to study is the bipartite extension of the Barab\'asi-Albert model~\cite{guillaume-2004-bipartite}. The results, shown in Fig.~\ref{fig:time_ba}, give a more complex picture than for the Erd\H{o}s-R\'enyi networks. The feedback-induced symmetry breaking persists, but other effects may also be at play, such as the influence of hubs (nodes with degree far exceeding the mean degree), which freeze the initiation rates at intermediate values. This is akin to the localization effects in epidemic models on scale-free networks~\cite{localization}. Following the same visual format and symmetric parameter grid, the extreme structural inequality of the Barab\'asi-Albert model causes severe variance and early divergence in individual agent trajectories. Crucially, in the high-drive regimes ($s \geq 0.5$), positive feedback induces symmetry breaking. Still, the rapid local saturation by hub individuals freezes macroscopic evolution early (cf.\ Ref.~\cite{holme_jo}), preventing the maximal gender disparity seen in the structurally less extreme Erd\H{o}s-R\'enyi model.

We have also analyzed the average final disparities $\Delta$ for the networked cases---analogous to Fig.~\ref{fig:final}---but these are so similar to the mean-field case that we do not show them here.

\section{Discussion}

We have analyzed a differential-equation-based model for initiation rates in online, heterosexual dating. Under a mean-field assumption, we can analyze the model analytically in detail. It shows complex behavior across three regions of the parameter space, each with distinct feedback patterns before saturation. In all cases, there is a saturation line of fixed points along which the sum of the initiation rates for the two genders equals the desired rate. Trajectories are attracted to this line from the interior, but once they reach it, Eq.~\eqref{eq:mf} has no motion along the line. The middle or end portions of the line should therefore be understood as possible final saturation outcomes selected by the earlier trajectory, not as points reached by later motion along the line. As the system approaches the saturation line, under fairly general conditions, feedback loops can drive a symmetry-breaking divergence in the initiation rates between the two genders. This divergence is much more pronounced when the system is confined to a sparse network, especially Erd\H{o}s-R\'enyi networks.

The networked version of the model connects the mechanism to platform design. The opportunity network $\mathbf{A}$ can be read as the outcome of a platform's recommendation and search features. These features change both the incoming activity $R_i$ that each user experiences and the load at which they saturate. In our simulations, amplification was strongest on sparse Erd\H{o}s--R\'enyi networks, whereas on Barab\'asi--Albert networks the mean gap was smaller because users around hubs saturated early. Personalization that confines users to small, local candidate sets could therefore make amplification more likely, while broader access would tend to reduce it, though not necessarily symmetrically. In the model, the symmetric state is also more robust when users' expectations of being contacted are modest relative to their contact budgets ($s<1/2$). Designs that restrict choice~\cite{halaburda2018competing} or which side may initiate~\cite{kanoria2021facilitating} have been analyzed for their effect on matching outcomes. Our results suggest that, regardless of the recommendation algorithm details, such choices can also affect gender disparities in initiation, even when both genders follow the same response rule.

The model deliberately leaves out several features of dating markets. Users do not value all contacts equally---who a candidate is, and how likely an approach is to be reciprocated, matter for contact and matching decisions~\cite{fisman2006gender,hitsch2010matching,bruch_extracting_2016}. Our preferred contact load is a reduced-form behavioral benchmark, not a claim that contact counts are users' ultimate objective. Deriving initiation from expected returns would require a choice rule and a model of how reciprocation depends on others' behavior, that is, a separate search-and-matching model. Our claim is accordingly narrower: the adjustment mechanism specified in our paper is sufficient for amplification. Whether incoming and outgoing contacts substitute for each other in practice, and whether approaches from more and less desirable candidates count alike, are empirical questions. Testing this requires longitudinal data on individual sending and receiving rates, together with measures of users' expected incoming contact relative to their available time.

Taken together, our core conclusion is that the widening of gender gaps in online dating does not require persistent external drivers. Three questions should be kept apart: where an initial asymmetry comes from, why it points in a particular direction, and why a small asymmetry becomes large. Our symmetric model answers only the third. Social norms about who should make the first move~\cite{eagly1987sex,finkel2009arbitrary}, gender differences in preferences, selectivity and reply rates~\cite{fisman2006gender,hitsch2010matching,bruch_aspirational_2018}, or features of the platform may create a modest initial difference or a modest asymmetry of $\beta$, and the feedback studied here then determines whether that difference fades or is magnified. In this view, endogenous dynamics and differences in norms or preferences are not competing explanations; they act at different stages of the same process. The individual's capacities and expectations may induce an instability that, through feedback from interactions across generations of users, creates an endogenous mechanism that increases gender disparity. More broadly, the debate on algorithmic fairness has emphasized advanced systems based on machine learning and artificial intelligence~\cite{dolata2022sociotechnical}. Our work shows that simple, seemingly innocuous design choices can introduce dynamic instabilities and lead to unintended consequences. We anticipate more future research in that direction.

\section*{Acknowledgements}
HT was supported by the Yunnan Provincial Department of Education Science Research Fund Project (2025Y0756) and the China Scholarship Council (202508530173). PH was supported by grant no.\ 376435 from the Research Council of Finland and JSPS KAKENHI Grant Number JP 25K01452.

\bibliographystyle{abbrv}
\bibliography{bib}

\end{document}